# A Blueprint for the Study of the Brain's Spatiotemporal Patterns

Emmanuelle Tognoli, Daniela Benites and J. A. Scott Kelso

Emmanuelle Tognoli is with the Center for Complex Systems and Brain Sciences, Florida Atlantic University, 777 Glades Rd., 33431 Boca Raton, Florida, USA (email: tognoli@ccs.fau.edu).

Daniela Benites is with the Center for Complex Systems and Brain Sciences, Florida Atlantic University, 777 Glades Rd., 33431 Boca Raton, Florida, USA and the Universidade Federal do Rio Grande do Sul, Porto Alegre, RS, Brazil (email: danibenites@gmail.com)

J. A. Scott Kelso is with the Center for Complex Systems and Brain Sciences, Florida Atlantic University, 777 Glades Rd., 33431 Boca Raton, Florida, USA and the Intelligent System Research Centre, University of Ulster, Magee campus, Northland Road, Derry, BT48 7JL, N. Ireland (email: kelso@ccs.fau.edu)

*Abstract*— The functioning of an organ such as the brain emerges from interactions between its constituent parts. Further, this interaction is not immutable in time but rather unfolds in a succession of patterns, thereby allowing the brain to adapt to constantly changing exterior and interior milieus. This calls for a framework able to study patterned spatiotemporal interactions between components of the brain. A theoretical and methodological framework is developed to study the brain's coordination dynamics. Here we present a toolset designed to decipher the continuous dynamics of electrophysiological data and its relation to (dys-) function. Understanding the spatiotemporal organization of brain patterns and their association with behavioral, cognitive and clinically-relevant variables is an important challenge for the fields of neuroscience and biologically-inspired engineering. It is hoped that such a comprehensive framework will shed light not only on human behavior and the human mind but also help in understanding the growing number of pathologies that are linked to disorders of brain connectivity.

## I. Introduction

The brain is a self-organized system that has been shown to operate in non-equilibrium regimes (Chialvo, 2010; Fuchs et al., 1992; Kelso et al., 1991; 1992; Tsuda, 2001). Dynamics is a chief aspect of brain function: dynamical descriptions are irreducible to typical time-aggregated quantities that are commonly employed to characterize brain activity (e.g. evoked potentials, spike rates, average spectra, static graphs of networks in electrophysiology). To understand how the brain functions, it is necessary to decipher its spatiotemporal organization and to do so with reference to continuous state variables describing the organism's behavior (Kelso, 1995; Kelso et al., 2013; Tognoli, 2008; Friston, 1997; Bressler & Kelso, 2001; Jung et al., 2001; Perez Velazquez & Frantseva, 2011). Only then can the rich context of external and internal variables be taken into account and the full complexity of the brain be acknowledged and described. Appropriate material for analysis under this goal is continuous recordings of neural activity, as illustrated here for example with continuous EEG (though other technologies such as magneto-encephalography (MEG), continuous functional Magnetic Resonance Imaging (fMRI), 2D and 3D Local Field Potential (LFP), multi-electrode array (MEA), functional Near-Infrared Spectroscopy (fNIRS) and other Optical Imaging techniques, etc. are also possible). It is sometimes argued: a) that continuous EEG suffers from insufficient signal-to-noise separation; and b) that efforts to work at the level of continuous dynamics have limited prospects except in cases of clear transitions such as those seen in sleep or epilepsy. As a consequence, data averaging is assumed to be necessary (Walter, 1964; Yeung et al., 2004). In our efforts to understand the continuous dynamics of the brain, we argue that it is complexity, rather than lack of systematic functional relevancy, that has obscured earlier efforts.

The ultimate goal of the present framework is to read the functional states of continuous brain dynamics as a comprehensible language, with spatiotemporal patterns at one end of the translation, and functional processes (behavioral, cognitive or clinical states) at the other. A method that allows deciphering of the spatiotemporal dynamics of the brain is significant to understand its functional organization, and may help explain complex aspects of human behavior that encompass perceiving and deciding, acting and knowing. Furthermore, such methods may be highly relevant to investigating the diseased brain, with the goal of (a) understanding intrinsic and distributed impacts of local dysfunctions and (b) studying brain disorders that are distributed in nature. Finally, a methodological framework is needed to test the effects of therapeutic interventions and in particular the pharmaco-dynamics of the brain. After a brief exposition of the theoretical framework of Coordination Dynamics, the following sections describe how to relate spatial, temporal and functional aspects of brain organization. Parts of the review will present methodological developments that have originated in the work of our team, whilst others will be more prospective and speculative, offering commentaries on how such new methods relate to existing techniques such as classification, source analysis, and functional inference. The conclusion addresses brain complexity and envisions obstacles and advances that the methodological framework may unlock.

## II. Theoretical Framework and Goals

The goal of Brain Coordination Dynamics is to understand how (the context-dependent) parts of the brain work together to produce the integrated experience that is behavior (in the largest sense of the word, including management of the organism's basal functions, perception, action, thoughts, emotion, memory, etc…). The methodological framework of Brain Coordination Dynamics is grounded in a theory of self-organization that provides models, predictions, and tools for the empirical study of brain function (Kelso, 1995; 2012; Kelso et al., 2013; Bressler & Kelso, 2001; Bressler & Tognoli, 2006; Tognoli & Kelso, 2009; 2014a). Brain Coordination Dynamics extends beyond the now well-accepted hypothesis that integration is accomplished by synchronization or phase locking (Tognoli & Kelso, 2014b), and it acknowledges and identifies more subtle and complex interactions between brain parts that reconcile complementary tendencies for collective as well as segregated activities (Kelso & Engstrøm, 2006). The generalized concept that is used to describe diverse modes of interaction between brain parts is that of coordination. Coordination may be defined as any dynamical regime in which information is flowing between brain parts, resulting in the emergence of a collective behavior that differs from the sum of the individual behaviors. It includes phase-locking or synchronization, which is probably the most sought-after mode of coupling between brain regions, but also other regimes in which meaningful interactions are preserved despite the disappearance of synchronization. Examples include relative coordination understood in terms of metastability, i.e., attractive states of the coordination dynamics replaced by more transient attracting tendencies (Kelso, 1995; 2012; Friston, 1997; Bressler & Kelso, 2001; Kelso & Tognoli, 2007; Werner, 2007; Rabinovich et al., 2008; Tognoli & Kelso, 2009; 2014a; Kelso et al., 2013; Bhowmik & Shanahan, 2013; Deco & Kringelbach, 2016). Once transient functional patterns and their modes of coordination are described (section III, Figure 1B), a further goal is to describe and explain transitions between them (section IV, Figure 1C), e.g., switching between attractors in a multistable regime in the presence of a perturbation, bifurcation observed when a control parameter changes or reorganization of dwells in a metastable regime. To achieve such a level of description, the succession of brain patterns needs to be fully characterized in continuous EEG. Finally, momentary brain states (section III) and their temporal organization (section IV) are related to inferred functions (section V, Figure 1D).

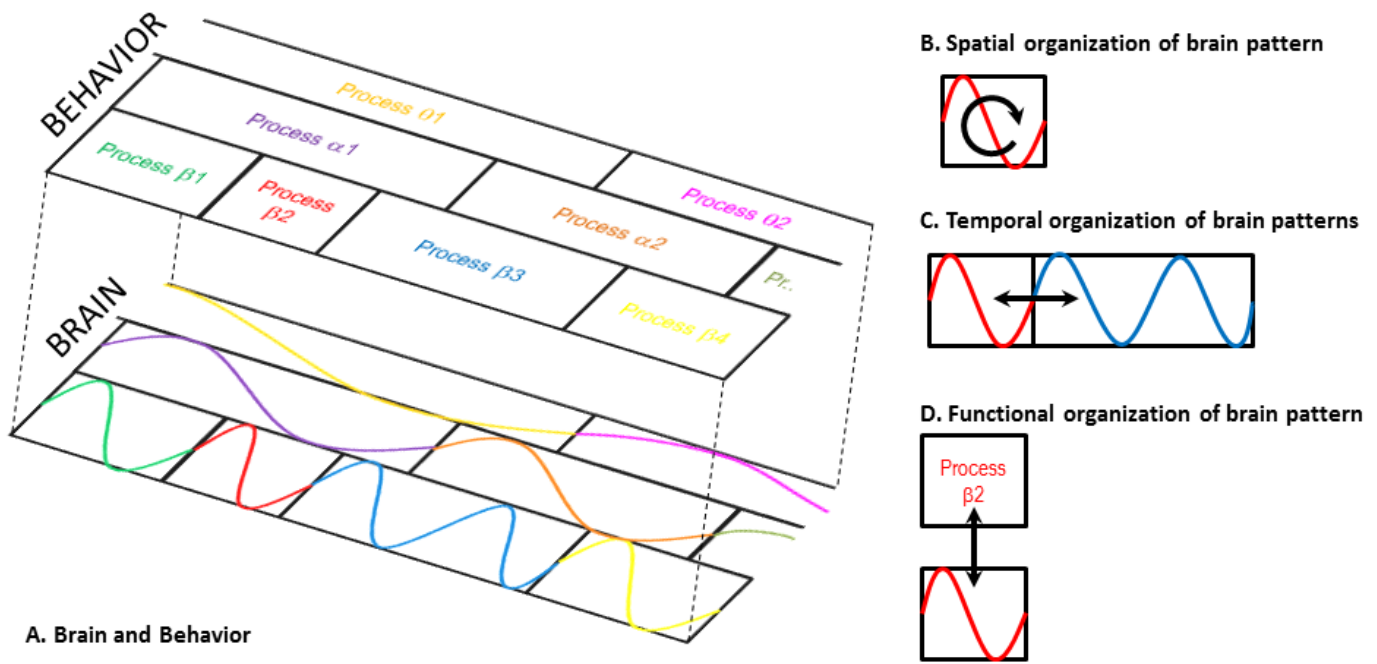

Fig. 1. Overview of the methodological framework of Brain Coordination Dynamics. The goal is to relate the spatiotemporal patterning of brain activity (illustrated with oscillatory patterns at different frequencies, A, lower frames) and behavioral/cognitive/clinical variables (A, upper frames). It is accomplished (B) by characterizing the spatial organization of brain patterns, that is, how neural ensembles interact over the duration of momentary brain states (intra-pattern organization as discussed in section III symbolized by the circular arrow); (C) by deciphering the temporal architectures of momentary brain patterns, that is, determining sequential rules of brain patterns (inter-pattern organization as discussed in section IV symbolized by the horizontal arrow connecting two successive patterns); and (D) by inferring relationship between states/transitions and (dys)functional processes (brain-behavior relationship as discussed in section V symbolized by the vertical arrow). For simplicity, here functional processes (e.g. processes of attention, memory, emotion, perception, etc…) are provisionally named in connection with their patterns' frequency band -greek letter- and serial temporal order -numeral suffix-).

## III. INTERNAL ORGANIZATION OF MOMENTARY BRAIN STATES

The fundamental units of neural dynamics are brief episodes of locally synchronous activity that rise and decay with a typical duration of just a few cycles during waking, rest and cognitive states (Tognoli & Kelso, 2009), likely as a result of spatiotemporally metastable organization at the level of the whole brain (Tognoli & Kelso, 2014a; Kelso, 2001; Kelso & Tognoli, 2006; 2007). The goal of this part of the methodological framework is to interpret the internal (spatial) organization of momentary brain patterns, namely, which neural populations are transiently recruited into collective behaviors and their dynamic interplay.

### A. Identifying oscillations: High resolution spectral analysis

Brain activity measured with EEG exhibits an oscillatory dynamics that spans frequencies from close to 0 (DC frequency) to hundreds of Hertz (Gobbelé et al., 1998; Canolty et al., 2006; Buzsaki & da Silva, 2012; Worrell et al., 2004; Bragin et al., 2010; Zijlmans et al.,

2012), covering time scales from sub-milliseconds to minutes and maybe more. Though the time-averaged spectrum (especially on long samples) suggests a continuous 1/f distribution with a few additional peaks (Pritchard, 1992), at any moment there are only few discrete frequencies at play (Roopun et al., 2008; Tognoli & Kelso, 2014a), with characteristic spatial, spectral and temporal footprints. Through the process of temporal averaging, such momentary oscillations lose their distinct appearance as peaks in the spectrum: as time accrues and other oscillations with similar spatial and spectral properties are added to the picture, the interval between peaks is filled in). To recover distinct component oscillations and their dynamics, it is useful to work in the temporal domain with continuous EEG (sections III.B-III.C). A broadband signal prevents a meaningful analysis of oscillatory phase dynamics: frequency bands need to be selected first which is where spectral analysis comes in.

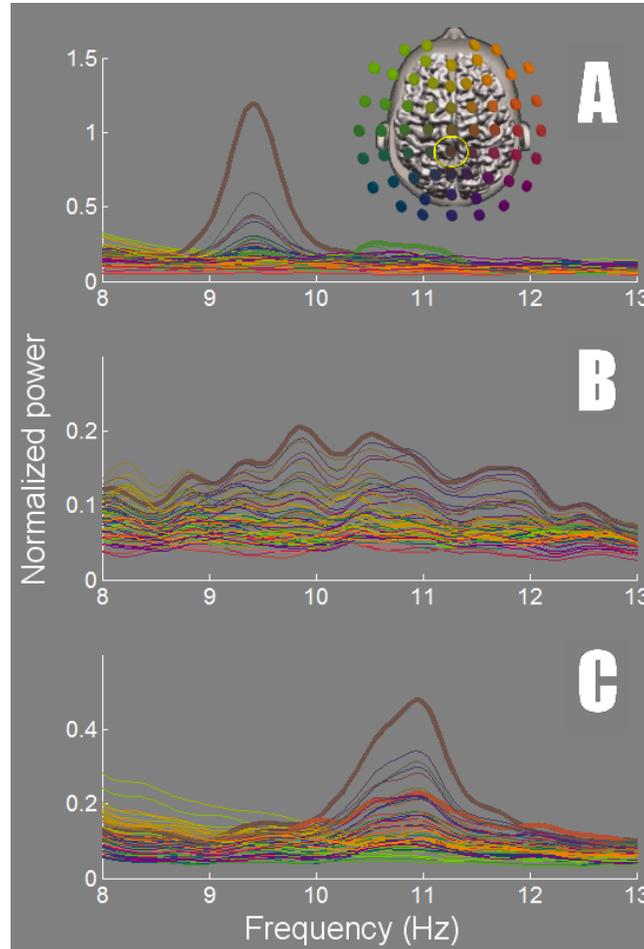

Fig. 2. Adapting frequency bands to each subject according to neuromarker idiosyncrasies. Sample colorimetric spectra in three subjects (A,B,C) during a task of action observation, showing neuromarker nu (thick brown line; Tognoli & Kelso, 2015), a rhythm from the 10Hz frequency band with peak power at electrode CPZ (reference placed at linked mastoids), spreading from 8.5 to 10.5Hz in subject (A), from 8.5 to 12.5Hz in subject (B) and from 10 to 12Hz in subject (C). Each spectrum is colorimetrically encoded according to its position on the scalp (see legend top right). The full alpha band (8-12.5Hz) fails to isolate nu properly in the subject shown in (A): this subject also harbors a left mu (green thick line peaking at 10.8Hz), whose task modulation may compromise the quantitative and functional interpretation of nu. It would also be unnecessarily wide for subject C. A narrower band would not manage to encompass all three subjects with the same pair of bandpass parameters. Therefore, filter parameters need to be adapted on a subject-by-subject basis

A long tradition of studying EEG oscillations in arbitrary frequency bands exists (e.g. alpha, beta, theta, delta, gamma, ripples and ultraslow oscillations). Although such frequency bands bear some relation to behavioral and cognitive processes (Basar et al., 1999; Von der Malsburg et al., 2010), their spectral boundaries are less than optimal. That is: (1) they are too wide to capture singular functional processes; and (2) if restricted to a meaningfully narrow band, they cease to properly capture inter-individual variability (meaning that the target functional process will be included in some subjects and excluded in others depending on how well aligned their frequency band is to the group norm).

The goal of high-resolution spectral analysis is to determine, on a subject-by-subject basis, which frequency bands are best for a target functional process (with its spatial, functional and spectral characteristics). Spectra can be estimated by a variety of means, for instance with the Fast Fourier Transform, Wavelet Transform, or Auto-Regressive modeling. Spectral peaks are either visible or buried depending on four qualities: their pattern's net amplitude, duration, a propensity to recur over the window of observation, and variability of their frequency footprint. If ample and lasting, recurring, and consistent in their frequency, oscillations may be revealed effortlessly from the raw spectrum (as in the case, e.g. of parieto-occipital alpha). If less manifest, oscillations can be emphasized by contrasts between experimental/clinical condition and a baseline, or in more subtle cases (brief oscillatory events with limited amplitude and recurrence) with selection of meaningful time-intervals based on a suitable theoretical model of the task/event.

Figure 2 show an example of EEG spectra in 3 different subjects, highlighting a neuromarker discovered by us in the 10Hz ("alpha") EEG band called nu which appears during social behavior (Tognoli & Kelso, 2015). A 2Hz-wide band would be adequate for subjects shown in Figure 2A and 2C, but they would differ in their spectral location (in low "alpha" 8.5-10.5Hz for subject A and high "alpha" 10-12Hz for subject C respectively). A 2Hz-wide band would not be suited for the subject shown in Figure 2B whose nu spans the 8.5-12.5Hz frequency band. Accordingly, a suitable analysis of those three subjects' nu oscillation and its relation to other neuromarkers requires adaptation of the frequency band to the idiosyncratic features of each subject.

The spectra's information of e.g. Figure 2 is critically enhanced with high spectral resolution. Our concern for high spectral resolution initially arose from our goal of examining neural oscillations' phase interactions (theory of phase-coupled oscillations, section II). Theoretically, oscillatory processes with interacting phases ought to have very similar frequencies to start with - so much so that coarse spectral resolution (especially when compounded with spatial proximity) makes the components indistinguishable from each other. In many contemporary studies, a typical frequency resolution is 1Hz. Our experience suggests that an enhanced (e.g. 0.1Hz) frequency interval (or bin size) provides a better spectral resolution to render closely related oscillations distinct (Tognoli et al., 2007; Tognoli & Kelso, 2013a). Such resolution requires either long continuous samples over 10 seconds (note conflicts with the temporal resolution called for by the typical time scales of brain patterns, in the order of 1-2 cycles, or e.g. about 1/10th of a second at 10Hz), or techniques to inflate spectral resolution such as zero-padding (preceded by suitable detrending and tapering to minimize artifacts).

A further commodity in the spectra's interpretation is provided by the colorimetric encoding of its spatial properties (Figure 2, legend top right). Breaking beyond the ceiling of 3D visualization, a technique was developed to study 4- or 5-dimensional spatiotemporal organization of EEG data (Tognoli et al., 2007; Tognoli & Kelso, 2009). This visualization superimposes all spectral series P(f) sampled in space (x,y) of e.g. the scalp surface, a four dimensional problem already. To render spatial organization, electrode location/proximity are converted to color/color similarity. This is realized by mapping spatial information (x,y) to a colorimetric model C, and then plotting each spectra P(f) with its color attribute C(x,y) as shown in the colorimetric legend of Figure 2, top right. In the resulting plot, spatial organization emerges as perceptually meaningful color organization (Figure 2A-C). The choice of color space is dictated by the geometry of the sensor spaces, in the case of EEG/MEG data, a half sphere or its projection on a disk are suitable. The color space is chosen so as to optimize perceptual abilities of the observer by selecting the largest color span available in the geometry. Successful visualization of the spatiotemporal patterning is based on perceptual color grouping (gestalt). For any observer with normal color vision, the spatio-spectral (or in forthcoming usages, spatio-temporal) patterning of the data is revealed quasi-instantaneously: it consists in identifying peaks with different colors (with color indicative of spatial organization).

### B. Selecting bandpass parameters to investigate phase dynamics

After meaningful frequencies are identified, they may be treated separately with the help of band-pass filters. This is done with the understanding that brain phenomena may occur in multiple frequencies (for instance, the wicket-shaped mu rhythm distributes energy both in the 10Hz "alpha/mu" and 20Hz "beta" frequency bands, Pfurtscheller et al., 1997; Hari, 2006), and that interactions between brain regions may be carried by coordination across frequencies (Kelso, 1995; Jensen & Colgin, 2007; Roopun et al., 2008). In such cases, filtering disperses meaningful information about source- or network-specific wave patterns in several datasets (the different frequency bands where information was distributed): ultimately, dispersed information will need proper methodology to get recouped.

The choice of frequency band(s) is a delicate matter but a key to successful analysis. Frequency selection should be data-driven whenever possible, i.e., based on the identification of oscillations (section III.A). The filter chosen must strike the right balance between too broad and too narrow. If the filter is too broad, reading the oscillations' spatiotemporal organization (phase dynamics) becomes difficult due to inclusion of too many unrelated frequencies. On the other hand, a narrow filter may clip oscillatory phenomena which fluctuate in frequency according to their diverse coordinative interactions with other oscillations. As the theory of coordination dynamics shows, 1:1 phase locking is possible for regions that exhibit different "natural" frequencies, for instance 10 and 12Hz. In the case of multifrequency coordination, other parametric regimes are also possible (e.g. Assisi, et al., 2005). Coupled neural ensembles will coordinate through a synchronized or metastable regime by compromising with each others' frequency preference (Kelso & Tognoli, 2007; Tognoli & Kelso, 2009). The choice of too narrow a filter may intermittently lose transient interactions between oscillatory neural ensembles. Colorimetric spectra (cf. Figure 2) reveal when a given peak departs from and returns to background. Such values of departure and return to background may be used for the filter cutoff. Furthermore, a soft roll-off (filter goes from passband to stopband with a gentle slope) is a useful way to minimize clipping of the oscillatory dynamics that occurs near the edges of the spectral process (left and right spectral margins of the oscillatory peak).

### C. Spatial organization (coordination dynamics within pattern)

After bandpass filtering, continuous analysis of brain dynamics reveals, band by band, the immediate nature of the brain's functional patterns (see samples in Figure 3). With remote (faster and slower than the filter's bandpass) frequencies removed, the patterns can usually be interpreted straightforwardly as the electric field from local areas, typically in numbers of one to just a few. The goal of this part of the analysis is to interpret how many local neural ensembles are involved in generating the signals, what their anatomical disposition is (radial or tangential or somewhere in between), and through which mechanisms they interact with each other.

Local neural ensembles provide an oscillatory pattern resembling that shown in Figure 3B (radial source, full dipole with phase aggregates of oscillations inphase and antiphase due to the instantaneous spatial pattern of their source's electric field) or Figure 3A (gyral source generating a truncated dipolar pattern, with a phase aggregate predominantly inphase. The antipole is absent because it would have appeared over the lower hemisphere of the head, where data are typically not collected). Identification of neural ensembles rests on the instantaneous correlation of the spatial pattern across many channels (a result of the dipolar electrical field, also referred to as volume conduction). As mentioned above, we call such instantaneous correlation a phase aggregate. Elementary patterns such as those of Figure 3A and 3B combine when more than one neural ensemble populates a frequency band (Figure 3C-G). The interactions can be weakest, with neural ensembles running at their own frequency and minimal phase attraction as in figure 3C (slow, green phase aggregate and faster blue phase aggregate); they can exhibit a metastable dynamics when neural ensembles running independently (early part of the pattern in Figure 3D, with the tan neural ensemble slower than the red one until they exhibit a phase-locking tendency over the second half of the pattern dynamics); or they can suggest phase locking out of phase (Figure 3E), inphase (Figure 3F) and antiphase (Figure 3G).

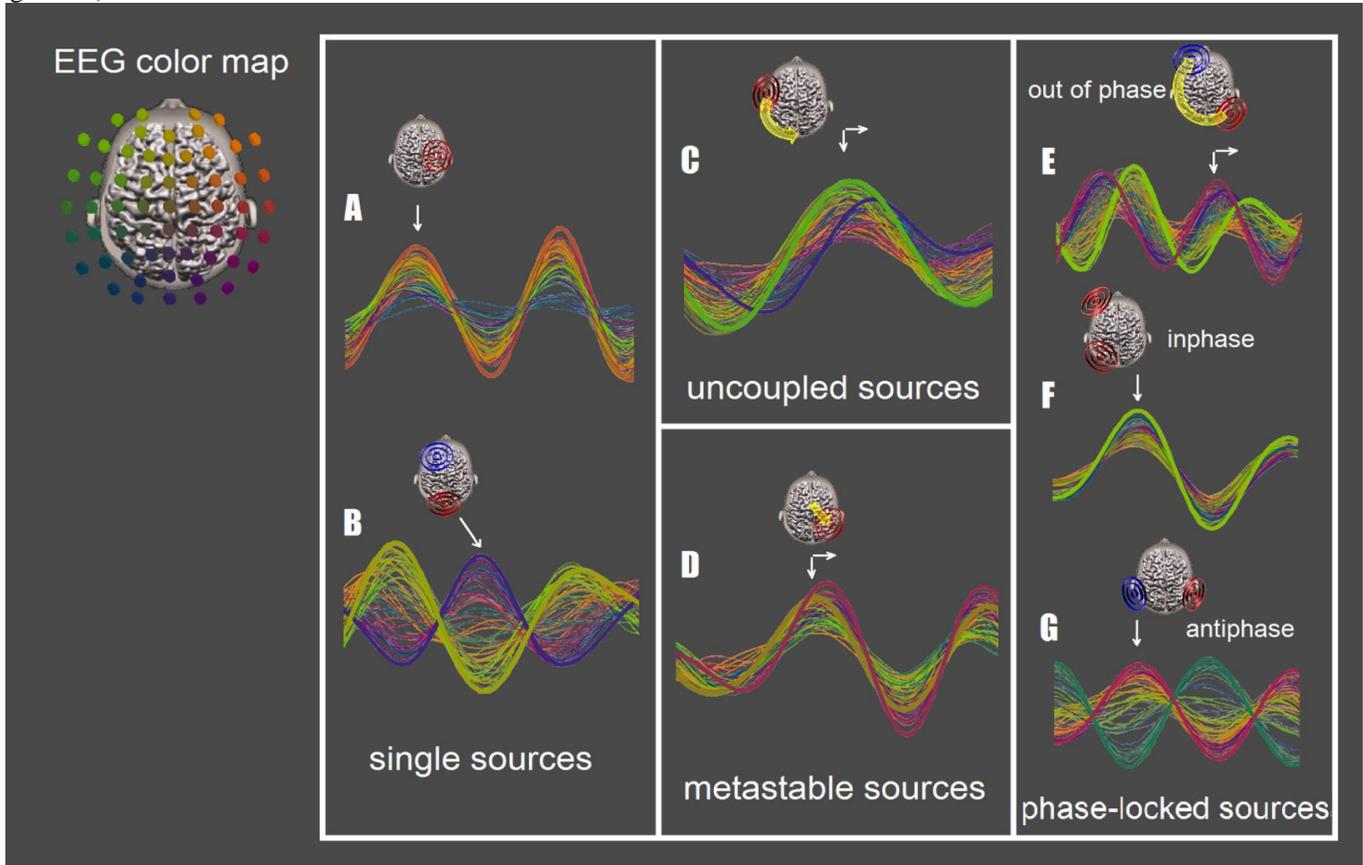

Fig. 3. Momentary spatiotemporal patterns with phase aggregates revealing discrete neural ensembles: (A) from a gyral source with its typical truncated dipole, here originating from a neural ensemble in the right temporal lobe; (B) from a sulcal source, with its scalp maxima in left occipital and left frontal regions, and a source placed in between them in the left rolandic area; (C) a pair of gyral sources without phase locking, (D) with metastable source dynamics, (E) with phase-locking out of phase, (F) inphase, and (G) antiphase. All time-series are colorimetrically encoded according to the legend shown on the right. Patterns' dynamic interpretation is based on a predictive forward work (Tognoli & Kelso, 2009).

The momentary patterns shown in Figure 3 correspond to a few neural ensembles that transiently enter into collective behavior as a result of mutual information exchange (section II). It is important to note that such transient regimes only persist for a few cycles, in line with the fast pace at which dynamical processes of cognition and behavior are presumed to occur. In the following, we examine the temporal organization of such brief patterns.

IV. TEMPORAL ORGANIZATION OF SUCCESSIVE PATTERNS

The previous section has provided guidelines to identify the brain's oscillatory patterns and what they are made of. This section completes that effort with a description of how patterns are organized in time. Time is an essential dimension here, since brain activity is non-stationary, and so are the functions associated with it (Tognoli & Kelso, 2014a). As we observe in most recordings, brain patterns may persist for one to just a few cycles and then give way to the next functional order. The temporal variation of brain patterns is obfuscated by averaging of either temporal segments (continuous EEG) or time points averaged across iterated trials (evoked analysis), both of which deliver falsely-recomposed pictures of brain dynamics (Tognoli, 2008). In the average picture, episodes of brain activity that never coexisted in the actual time course of brain patterns suddenly appear together. Also, the collapse of the dynamics diminishes the ability to tease apart brain mechanisms (e.g. transient functional networks) through a theory-guided interpretation of continuous neural activity.

Theories have suggested that brain patterns might be articulated into sequences (Lehmann et al., 1987; Kropff & Treves, 2007; Rabinovich et al., 2008; Tognoli & Kelso, 2014a; Tsuda, 2015), and if such is the case, it might be possible to detect statistical traces of a temporal order between patterns. Within a framework that preserves the genuine dynamics of brain activity, we can start to look into this question of serial organization: are there patterns Y that preferentially follow (or avoid) other patterns X? Does this characteristic temporal order lead to functional modulation in behavioral performance, e.g., better functional outcomes for certain sequences than for others or even bifurcations between healthy and disease states? Are sequences of patterns part of a syntax for the working brain? To address such issues, our analysis starts with the identification of pattern boundaries (segmentation of continuous, band-filtered EEG), followed by pattern classification and investigation of the classified patterns' temporal dependencies.

*A. Segmenting continuous EEG*

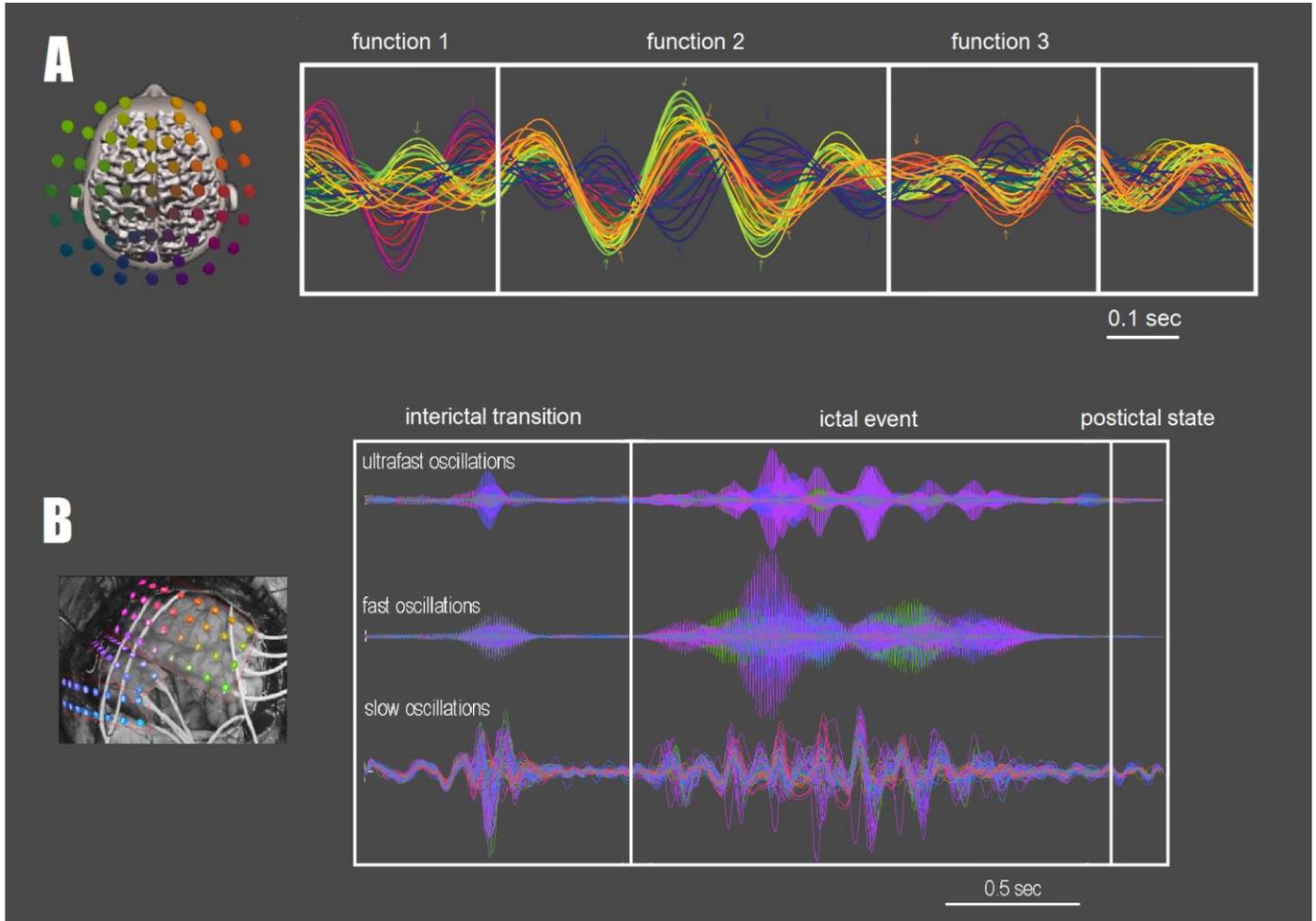

Fig. 4. Continuous segmentation in A reveals brain-behavior dynamics in a signal filtered around 10Hz recorded during social coordination, with the first pattern showing magenta and green aggregates (first box) replaced by a pattern with green, orange and blue aggregates (second box), orange and purple in the third box, and the phase scattering of neural populations in the 4th box. Each pattern is hypothesized to play a functional role. In B, a seizure is presented to illustrate the multiscale dynamics across frequency bands. Note that as the seizure spreads, patterns change quickly in the fastest frequencies (successive bursts with different colors, top graph), but much more slowly in the lowest frequencies (steady pattern occupies most of the slow oscillation band in lower graph).

To clearly divide successive patterns of brain activity, a qualitative change of dynamics needs to be recorded. Such qualitative change may arise when neural ensembles enter or leave the ongoing coordination pattern. This occurs when the system is in an attractor regime, (a) as a result of a transition between multistable states due to external input or noise, or (b) as a bifurcation when the system undergoes a transformation in its dynamic landscape. Qualitative changes may also occur while the system is in a stationary transient regime sans attractor: (c) arising at the edge of metastability's "escape" and "dwell" times (Kelso & Tognoli, 2007; Kelso, 2012; Tognoli & Kelso, 2014a). Figure 4 displays examples of segmented EEG. It shows typical phase aggregates as described in section III.C. As long as the spatiotemporal organization continues unaltered, we can assume that the system is in the same coordination pattern. If a change is observed, for instance one phase aggregate disappears (Figure 4A, 1st box), shifts spatially, or the coordination dynamics of the ongoing pattern is reorganized (Figure 4A, 4th box), then we can assume a transition has occurred. Such transitions can be marked reliably by visual inspection (Benites et al., 2010; *in press*) or they can be estimated with algorithms that track switches in the spatiotemporal organization (Fuchs et al., 2010). The output of this analysis is a chain of transient spatiotemporal patterns dissected from the continuous EEG. Each pattern expresses the coordinated behavior of a discrete set of brain regions and relates to a transient behavioral or clinically-relevant organization. Figure 4B presents the electrocorticogram (ECoG) record of an epileptic seizure (after data from Nadasdy et al., 2012). It shows that the dynamic patterning is specific to the oscillations' frequency band, with more frequent pattern changes occurring in the fastest frequencies (successions of bursts of different colors in top graph as seizure spreads) and gradually slower turnover of patterns at lower frequency bands (Figure 4B, middle and lower graphs). In behavior and cognition, we hypothesize that patterns persist for as long as it is necessary for information to be exchanged within originating brain areas. They may persist longer in functions requiring sustained dynamics (e.g. memory, attention). In clinical contexts, patterns might last as long as certain neurochemical and metabolic transactions supporting them persist (e.g. epilepsy).

B.  *Classifying patterns, option 1: at the scalp level*

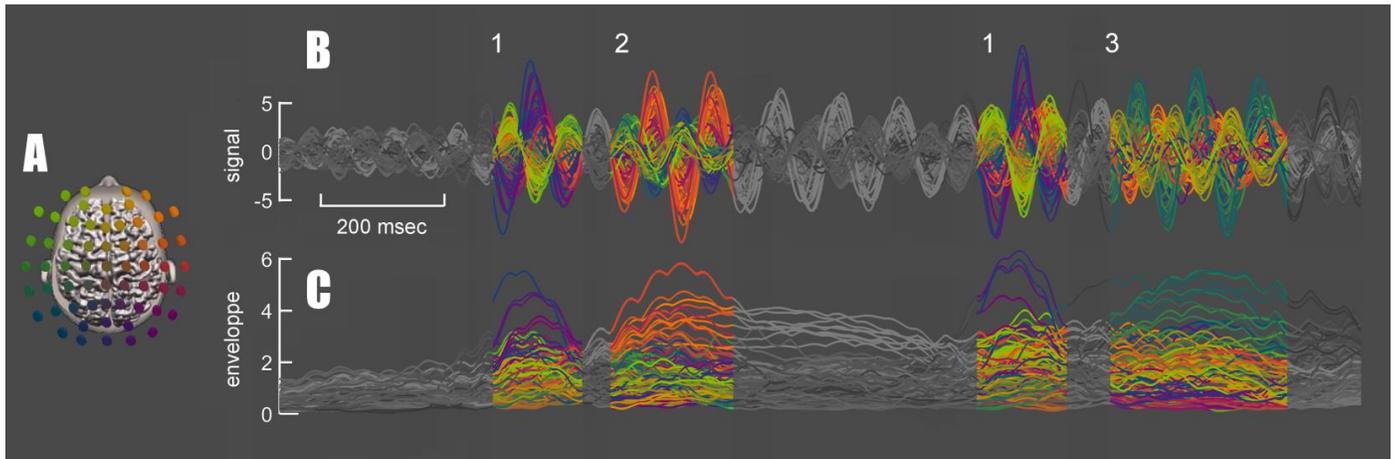

Fig. 5. Patterns can be classified to identify recurrence of the same brain network, as seen in this example of rest state EEG, where four patterns are highlighted (top, band passed filtered oscillations, and bottom, their envelope -a measure of instantaneous power-) with colorimetric encoding shown on the right, and first and third pattern belonging to the same class, with co-existing blue and orange phase aggregates.

Once patterns are divided into meaningful segments, efforts are undertaken to pool them according to similar spatiotemporal characteristics (see e.g. Figure 5). The goal is to recognize, from the diversity of patterns, recurring brain networks contributing to the repeated occurrence of functional or clinical processes. The components of transient brain patterns, namely local neural populations, yield variations in their scalp appearance in different network contexts. For instance a local population A will generate two different patterns (phase aggregates with different scalp topographical and spectral properties) when occurring along with local populations B or C. This is because different associations will alter spatial, frequency, phase and amplitude properties as a result of their electric fields' combination (Tognoli & Kelso, 2009). The compositional outcome of the entire network is immune to this variation (phase aggregates are consistent within the same network context), thereby allowing meaningful classifications at the level of the whole pattern. The classifier might be any of a number of methods such as Artificial Neural Networks, Support Vector Machine, Discriminant Analysis, Common Spatial Pattern analysis (e.g. Lotte et al., 2007; Schindler et al., 2007). The attributes that are essential for classification include phase aggregates' topography, frequency (measured at the lead electrode for minimal bias, see Tognoli & Kelso, 2009), and in the usual case when multiple phase aggregates are present, the type of coordination they exhibit (Figure 3) along with their phase relationship. Note that this is in contrast to approaches that blindly feed raw data to a classifier in the hope that some patterns will emerge. This is because multichannel data feed to classifiers carries all the spatial redundancy of brain sources' electrical fields. A source quasi-instantaneously distributes its signal across the whole scalp surface (see, e.g., figure 3A from Tognoli & Kelso, 2009). Spatial ubiquity of the electrical fields is a biasing circumstance for the classifier: it emphasizes random aspects of the pattern depending on source number, orientation, and other factors.

C.  *Classifying patterns, option 2: at the source level*

Alternatively, and contingent on the availability of suitable models of the subject's head and brain, the segmented patterns of section IV.A are ideal candidates for source estimation, thereby leading to better interpretation of the underlying brain dynamics. Sources may be estimated first, and then the sources' configuration and dynamics are subjected to classification. We stress that segmented epochs of brain patterns are ideal for source estimation. Theoretically, source estimation is ill-conceived when conducted on average signals such as Evoked Potentials (though in practice and heuristically, it often works sufficiently well, especially for radial sources). This theoretical limitation is because Evoked Potentials are a mathematical reconstruction that fundamentally departs from the continuous dynamics of local electrical fields responsible for producing them (Tognoli, 2008). Alternatively, epochs sampled from continuous EEG can be used for source estimation. If source estimation is applied to fixed-duration epochs of continuous EEG (a common procedure), it is obvious from the spatiotemporal dynamics of Figure 4 that dissident components (components that do not belong to the same pattern) might be mixed according to the vagaries of epoch cutting. This mixing comes at the cost of source estimation: the algorithm is forced to explain a pattern that never was and as a result, it will likely choose an incorrect solution. Thus, source estimation applied to fixed-duration epochs defeats the hope of capturing transient brain organization, with inverse solutions degrading for shorter epochs. Longer epochs might remain more resilient, but they don't address the dynamics at a proper temporal scale, and furthermore, they capture some "default" activity rather than more relevant moment-to-moment changing patterns. The issue is solved when homogeneous spatiotemporal patterns (the outcome of segmentation, section IV.A) are subjected to source estimation. The source estimation algorithm can be fed data that contains all the required information, and only the required information about a transient brain pattern. The outcome of such a process is the specification of local sources and their dynamics. The classifier can then be applied either on variables specifying the individual sources (location, amplitude, frequency and phase) or on their combination in a pattern.

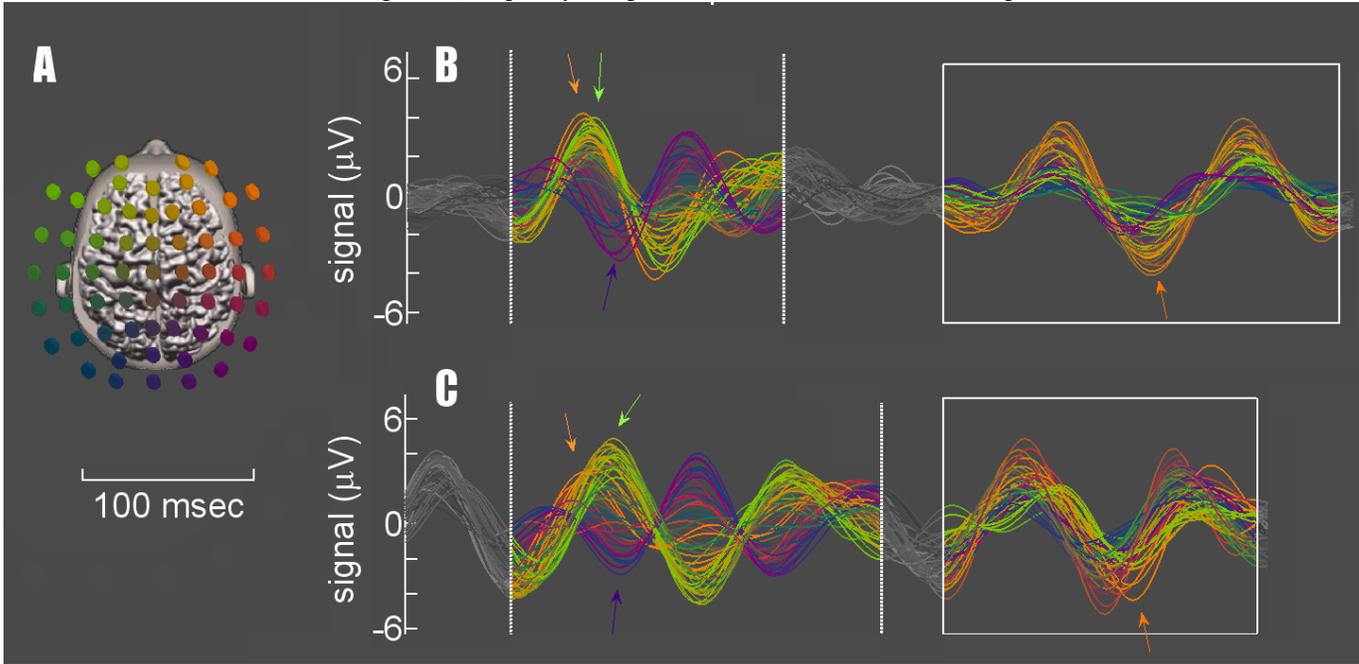

Fig. 6. Example of two EEG sequences with pattern repetition (B-C), from two subjects engaged in social sensorimotor coordination. First pattern has three phase aggregates, in orange, green and indigo, second is dominated by one phase aggregate, a dark orange truncated dipole with the zero potential in the left centro-parietal region (teal color). Note inter-individual difference in pattern frequency (first pattern is faster in subject B, slower in C; and vice versa for the second pattern). Also note additional activity in the second pattern of subject C (other phase aggregate in a darker tone of orange). Repeated sequences are observed both intra- and inter-individually, suggesting their relevance to task-dependent or task-independent patterns of behavior.

*D. Temporal dependencies between patterns*

The ultimate goal of a spatiotemporal analysis of brain activity is to peek into the mechanisms of brain function, viewed here as dynamical descriptions that link behavioral, cognitive or clinical states on one hand, and their associated transient neural activities on the other. Temporal organization has been advanced on the functional/clinical side (such as serial models of mental operations in behavioral and cognitive science, or stage identification in the case of diseases, see section V) but it has remained more elusive with respect to underlying brain patterns--probably owing to the dearth of dynamical approaches. The goal of this part of the framework is a quantitative description of the temporal organization of brain patterns: which tend to precede or follow which? Temporal organization has been theorized in terms of metastability (Kelso, 1995; Tognoli & Kelso, 2013b; 2014a); phase transitions (Freeman, 2006; Kelso, 1995; Kelso, et al., 1991; 1992; 2012); chaotic itinerancy (Tsuda, 2015; Kelso & Fuchs, 1995) or heteroclinic channels (Rabinovich et al., 2008) amongst others. We hypothesize that the brain has too much complexity and degeneracy for successive realizations of the same circumstances to yield the exact same sequence of patterns. But if brain patterns are to play a role in function, they most likely carry a definite temporal order (see example in Figure 6). To our knowledge, no specific study has established the scope of this dynamic reproducibility in a quantitative manner. Therefore, provided our currently limited understanding of how the brain works, the study of its patterns' temporal organization has to be a statistical one. Once patterns are classified (section IV.B-IV.C), the EEG can be rewritten as a sparse series of transient patterns that follow each other, and for each pattern type, the frequency of its preceding and following pattern types can be counted. Those sequential events can be tested against the null hypothesis that temporal organization is random, that is, successive patterns occur no more or no less frequently than chance provided their total probability of occurrence (Benites et al., 2010). Those sequences which are statistically more likely to appear than by chance, are of greatest interest for a formal description of the dynamics of brain and behavior. They are seeds for objective understanding of the brain mechanisms supporting behavioral, cognitive and clinical processes.

V. FUNCTIONAL ORGANIZATION

The last part of the challenge described in Figure 1 entails a dynamical description of the behavioral, cognitive or clinical processes that are manifested on the functional side of the brain~behavior relationship: what function is served from moment to moment by the brain's transient patterns. This is accomplished by a study of statistical associations with, e.g. chi square statistics. Ideally, the goal of identifying linkages between functional processes and neural patterns would be guided by complete dynamical models of those functional processes – thus leaving us with the simpler task of matching the two. First, we discuss the extent and limits of this ideal circumstance (see also Bechtel, 2004, for related account). Next, we identify strategies to decipher the dynamics of functional processes and their matching with brain patterns.

## A. Canonical models of brain function

Currently no functional model exists that extensively follows the scheme shown in Figure 1. The latter represents a long-term goal of the framework of Brain Coordination Dynamics, a puzzle to be completed piece by piece. The template of Figure 1 hypothesizes several concurrent processes whose different durations match their neural patterns' typical timescales (*qua* frequency band). Many models are available of behavioral, cognitive and clinical tasks or circumstances. Yet despite an era of cognitive science that brought attention to the brain's capabilities for "parallel" processing (e.g. McClelland, Rumelhart et al., 1986; Mesulam, 1990), the models typically do not elaborate multiple functional processes running in parallel on their own time scales; instead they tend to focus on a single thread, with only one process at a time (but see e.g. Canolty & Knight, 2010 for renewed interest and developments supporting multiscale neurofunctional processes). Even though most models are serial, their hypothesized functional processes can be a useful starting-point to elucidate the neurofunctional organization of the brain.

To create dynamical models of brain function, and identify 'when' processes occur, of the three domains of investigation considered in this work, behavior is often simpler to deal with since it is readily amenable to continuous measurement of state variables over time (Tognoli, 2008). Cognition is more elusive, since it deals with a covert phenomenology, one that provides only indirect and discontinuous traces of essential processes. Indeed, the enormous growth of the Neuroscience today is largely based on the confines of a Cognitive paradigm, with Neuroscience built on the premises that neural descriptions of processes will fill a gap left unresolved by Psychological and Cognitive Science. A chief tenet of this methodology is that whenever a brain area or a network is registered in a certain circumstance, its observation calls in a vast amount of prior data regarding the purpose that this area/network plays in other studies. Interpretative work tries to integrate and refine the area/network's functional meaning, given the intentions behind present and past studies and experimental manipulations. Even when continuous quantification of functional processes is available, it remains a challenge to identify which variables are meaningful, and which are not.

A rich history of experimental studies is available for all three domains considered here -behavioral, cognitive and clinical- that manipulates one factor (e.g. memory load, clinical and neurochemical state, perceptual features, action goal), and identifies certain brain parts or their indirect manifestations (as in EEG) responsible for that factor's role. To place meaningful processes in a temporal dimension, there is also the approach from Coordination Dynamics that uses phase transitions as a basis for meaningful, temporally localizable phenomena presiding over the (re)organization of the system (Kelso, 1995; Kelso et al., 2013; see Benites et al., 2010 for exemplary application). Canonical models of the dynamics of the brain's functional processes are still an industry in the making, with some partial answers on the 'what' (relevant variables) and 'when' (dynamics), though oftentimes still lacking an association between the two. Progress will depend on efforts aimed at a fuller perspective of the brain's multiple simultaneous functional processes. Due to their complexity, such functional processes may well be unique to each iteration and subject. It thus becomes useful to consider replacing canonical models (material for confirmatory analysis of theoretical models of the brain' functional dynamics) with descriptions of individual instances of functional dynamics. We discuss this more exploratory approach next, also one when variation in the evolution of the system, both useful and detrimental, can be systematically characterized.

## B. Individualized models of brain function

Instead of working from a functional model to the data, the opposite -exploratory- perspective can be pursued: from the data to the model. This asks the question: "are there observed circumstances O under which a brain pattern of interest is more (less) frequent than chance (or more/less ample, lasting, or varying along any relevant characteristics of the pattern)?" Neither in our perspective nor in the previously discussed paradigms of Experimental and Cognitive Neuroscience (section V.A), is there a certainty that the core variables explaining a brain pattern P are the ones being manipulated. An observed variable O that reveals an association with a brain pattern P might more commonly have another statistical dependency with the hidden (but truly explanatory) variable E, and we might observe an association between brain pattern P and function O that is mediated by the association E-O. In this case, the desirable interpretation f (E) is unreachable, and any interpretation f (O) would be an incorrect byproduct of the association E-O. To properly represent the complexity of the brain's neurofunctional organization and enhance the odds that explanatory variables belong to the analysis, a dense-enough set of descriptors (nominal or quantitative variables, both dependent and independent) may be used to characterize the experimental context of each trial/moment. For instance, in a work to identify the neurofunctional patterns typical of several types of transitions between social and individual behavior, Benites et al., 2010 used a combinatorial set of 288 descriptors, themselves based upon the association of 7 modal independent and dependent variables. Pattern properties (e.g. frequency of occurrence) were compared across descriptors, and tentative interpretation of their meaning was based on the strongest significant associations (assuming that the greatest variance is explained by true explanatory variables E, and lesser variance emerges in covariates O). Note that some multivariate modeling (e.g. structural equation modeling, MANOVA) could complement this approach.

As with the Cognitive Neuroscience paradigm, our functional analysis allows to start attributing meaning to some brain patterns (with the aforementioned cautions about covariates in mind), and it fills the picture of Figure 1D with regard to functional inference. In contrast to much of Cognitive Neuroscience though, such a dynamical framework opens up the analysis of inter-individual and inter-trial variation in neurofunctional organization. A study delivers not one canonical set of processes for the completion of a task or for the evolution of a clinical episode (section V.A) but as many as were intended experimentally by the repetition of identical trials and similar classes of subjects (section V.B), within session (a most common practice for cognition and behavior) and across sessions, for instance as part of a long term monitoring of neuro-functional reorganization in disease or in development. The benefit of this multiple approach is the possibility to describe degeneracy in the brain and behavior (Kelso & Tuller, 1984; Edelman & Gally, 2001, Dodel et al., 2013). Degeneracy here means the fulfillment of the same function by a different circuitry or dynamics and hence a different organization of brain patterns.

## VI. Summary and outlook

The brain seems to operate by altering the patterning of its activity at specific places and times via a vast web of partially specialized neural populations. This self-organizing process can apparently be guided from within or under the influence of external nudges by the body and the environment. A true dynamical framework to analyze the brain's continuous signals is still a masterpiece in the making, rendered difficult by the complicated interplay of neural activity in space and time (Elbert & Keil, 2000; Dalal et al., 2008; Tognoli & Kelso, 2014b). The present framework exploits our ability to visualize and interpret brain signals in 4 or 5 dimensions of space-time (Tognoli & Kelso, 2010). At the macroscopic levels of EEG (and MEG and other collective signals from the brain's vastly numerous neurons), the present framework has revealed that human brain patterns are short, sparse, intermittent and coordinated temporally. A few local areas matter at any given time, with their composition changing swiftly from moment to moment. The fact that the brain exhibits patterned activity at all suggests that it lives in a reasonably low dimensional state space—though its dynamics may be quite complicated (Kelso, 2012). The net outcome of our experimental observations is a series of transient states where a few local neural populations interact at any moment through phase locking or metastability in each of a few frequency bands (Tognoli & Kelso, 2009, 2014a; section III). Accordingly, the combinatorial complexity of the brain is manifested, not instantaneously, but in the succession of its momentary patterns, with many more oscillations relevant to brain function than suggested by the historical compilation of a handful of brain rhythms (delta, theta, alpha, beta, gamma…). At the same time, with the patterns cleaved from one another in a meaningful manner through the process of spatiotemporal sequencing (section IV.A-D), an enhanced grasp of the brain's true coordination dynamics is achieved (Tognoli & Kelso, 2009), and better prospects arise to accurately describe the location and dynamics of cortical sources (section IV.C).

We hypothesize that brain performance depends in part upon variation on the path leading from patterns to patterns within individual (under variation of mental and behavioral states), as well as between individuals (as a function of their skills). We anticipate that analyzing the properties of brain patterns and their temporal organization under variation of task performance will help define the multiple paths available for the brain to work, and their relative benefits under different constraints (resources, contexts) and goals (performance). Such an effort addresses the exceptional complexity and degeneracy of brain structure, function and behavior (Edelman & Gally, 2001; Bressler & Tognoli, 2006; Kelso & Tuller, 1984; Tognoli & Kelso, 2014b) that has hardly begun to be quantified. Description at a neurophysiological level is not sufficient in and of itself. Understanding depends on what is accomplished functionally – and we hypothesize that each spatiotemporal pattern realizes a piece of function (or dysfunction). Under this hypothesis, our goal is to attain a translational understanding linking spatiotemporal patterns and the functional processes that such momentary states of brain organization support, so that either level of description is substitutable by the other. Section V highlighted the challenges and opportunities of such an endeavor. The incompleteness of formal dynamical descriptions of behavioral/cognitive/clinical processes, viz., a lack of theories of tasks, was raised as a difficulty, but it also provided an opportunity for our framework to advance both in terms of multiscale canonical models of brain function (with the largest gains expected for processes at certain time scales that have eluded past phenomenological descriptions, e.g. very fast frequencies), and in terms of specific characterization of individual instances of task performance. The latter part closes the loop on degeneracy and relates neurophysiological and functional idiosyncrasies. In our framework, idiosyncrasies are a healthy sign of brain complexity and adaptability: they are taken as opportunities to understand which processes are available under a given set of circumstances, and hence may lead to better training and remediation strategies in both health and disease.

Acknowledgments:

The authors are grateful for the support of grants from the National Institute of Mental Health (MH080838), the National Institute for Biomedical Imaging and Bioengineering (NIBIB EB025819), the National Science Foundation (BCS0826897), the US Office of Naval Research (N000140910527), the Chaire d'Excellence Pierre de Fermat and the Davimos Family Endowment for Excellence in Science.

## References

[1] Chialvo, D. R. (2010). Emergent complex neural dynamics. Nature physics, 6(10), 744-750.
[2] Fuchs, A., Kelso, J.A.S., Haken, H. (1992). Phase transitions in the human brain: spatial mode dynamics. International Journal of Bifurcation and Chaos, 2(4): 917-939.
[3] Kelso, J.A.S., Bressler, S.L., Buchanan, S., DeGuzman, G.C., Ding, M., Fuchs, A. Holroyd, T. (1991). Cooperative and critical phenomena in the human brain revealed by multiple SQUIDS. In D. Duke & W. Pritchard, (Eds.), Measuring Chaos in the Human Brain, pp. 97 112. World Scientific, New Jersey.
[4] Kelso, J.A.S., Bressler, S.L., Buchanan, S., DeGuzman, G.C., Ding, M., Fuchs, A. Holroyd, T. (1992). A phase transition in human brain and behavior. Physics Letters A, 169, 134 144.
[5] Tsuda, I., (2001). Toward an interpretation of dynamic neural activity in terms of chaotic dynamical systems. Behav Brain Sci., 24(5): 793-810.
[6] Kelso, J.A.S. (1995). Dynamic Patterns: The self-organization of brain and behavior. Cambridge, MIT Press.
[7] Kelso, J. A. S., Dumas, G., Tognoli, E. (2013). Outline of a general theory of behavior and brain coordination. Neural Networks, 37, 120-131.
[8] Tognoli, E. (2008). EEG coordination dynamics: neuromarkers of social coordination. In Coordination: Neural, Behavioral and Social Dynamics (pp. 309-323). Springer Berlin Heidelberg.
[9] Friston, K. J. (1997). Transients, metastability, and neuronal dynamics. Neuroimage, 5(2): 164-171.
[10] Bressler, S. L., Kelso, J. A. S. (2001). Cortical coordination dynamics and cognition. Trends in Cognitive Sciences 5 (1): 26-36.

[11] Jung, T. P., Makeig, S., McKeown, M. J., Bell, A. J., Lee, T. W., Sejnowski, T. J. (2001). Imaging brain dynamics using independent component analysis. Proceedings of the IEEE, 89(7), 1107-1122.
[12] Perez-Velazquez, J. L., Frantseva, M. (2011). The Brain-behavior Continuum: The Subtle Transition Between Sanity and Insanity. World Scientific Publishing Co., Singapore.
[13] Walter, W. G. (1964). Slow potential waves in the human brain associated with expectancy, attention and decision. European Archives of Psychiatry and Clinical Neuroscience, 206(3), 309-322.
[14] Yeung, N., Bogacz, R., Holroyd, C. B., Cohen, J. D. (2004). Detection of synchronized oscillations in the electroencephalogram: an evaluation of methods. Psychophysiology, 41(6), 822-832.
[15] Kelso, J.A.S. (2012). Multistability and metastability: Understanding dynamic coordination in the brain. Phil. Trans. Royal Society B, 367, 906-918.
[16] Bressler, S.L., Tognoli, E. (2006). Operational principles of neurocognitive networks. International Journal of Psychophysiology, 60: 139-148.
[17] Tognoli, E., Kelso, J. A. S. (2009). Brain coordination dynamics: true and false faces of phase synchrony and metastability. Progress in Neurobiology, 87(1), 31-40.
[18] Tognoli, E., Kelso, J. A. S. (2014a). The metastable brain. Neuron, 81(1), 35-48.
[19] Tognoli, E., Kelso, J. A. S. (2014b). Enlarging the scope: grasping brain complexity. Frontiers in systems neuroscience, 8: 122.
[20] Kelso, J. A. S., Engstrøm, D. A. (2006). The complementary nature. The MIT Press.
[21] Kelso, J.A.S., Tognoli, E. (2007) Toward a complementary neuroscience: metastable coordination dynamics of the brain. In R. Kozma and L. Perlovsky (Eds.) Neurodynamics of Cognition and Consciousness. Springer, Heidelberg, pp.39-60.
[22] Werner, G. (2007). Metastability, criticality and phase transitions in brain and its models. Biosystems, 90(2), 496-508.
[23] Rabinovich, M.I., Huerta, R., Varona, P., Afraimovich, V.S. (2008a) Transient cognitive dynamics, metastability, and decision making. PLoS Comput Biol, 4(5): e1000072.
[24] Bhowmik, D., Shanahan, M. (2013). Metastability and inter-band frequency modulation in networks of oscillating spiking neuron populations. PloS one, 8(4), e62234.
[25] Deco, G., Kringelbach, M. L. (2016). Metastability and coherence: extending the communication through coherence hypothesis using a whole-brain computational perspective. Trends in neurosciences, 39(3), 125-135.
[26] Kelso. J.A.S. (2001). Metastable coordination dynamics of brain and behavior. Brain and Neural Networks (Japan) 8, 125-130.
[27] Kelso, J.A.S., Tognoli, E. (2006). Metastability in the brain. Proceedings of the International Joint Conference on Neural Networks, Vancouver, pp.755-760.
[28] Gobbelé, R., Buchner, H., Curio, G. (1998). High-frequency (600 Hz) SEP activities originating in the subcortical and cortical human somatosensory system. Electroencephalography and Clinical Neurophysiology, Evoked Potentials Section, 108(2), 182-189.
[29] Canolty, R. T., Edwards, E., Dalal, S. S., Soltani, M., Nagarajan, S. S., Kirsch, H. E., Berger, M. S., Barbaro, N. M., Knight, R. T. (2006). High gamma power is phase-locked to theta oscillations in human neocortex. science, 313(5793), 1626-1628.
[30] Buzsáki, G., da Silva, F. L. (2012). High frequency oscillations in the intact brain. Progress in neurobiology, 98(3), 241-249.
[31] Worrell, G. A., Parish, L., Cranstoun, S. D., Jonas, R., Baltuch, G., Litt, B. (2004). High-frequency oscillations and seizure generation in neocortical epilepsy. Brain, 127(7), 1496-1506.
[32] Bragin, A., Engel Jr, J., Staba, R. J. (2010). High-frequency oscillations in epileptic brain. Current opinion in neurology, 23(2), 151.
[33] Zijlmans, M., Jiruska, P., Zelmann, R., Leijten, F. S., Jefferys, J. G., Gotman, J. (2012). High-frequency oscillations as a new biomarker in epilepsy. Annals of neurology, 71(2), 169-178.
[34] Pritchard, W.S. (1992). The brain in fractal time: 1/f-like power spectrum scaling of the human electroencephalogram. Int. J. Neurosci. 66(1-2): 119-129.
[35] Roopun, A. K., Kramer, M. A., Carracedo, L. M., Kaiser, M., Davies, C. H., Traub, R. D., Kopell, N.J., Whittington, M. A. (2008). Temporal interactions between cortical rhythms. Frontiers in neuroscience, 2(2), 145.
[36] Başar, E., Başar-Eroğlu, C., Karakaş, S., Schürmann, M. (1999). Are cognitive processes manifested in event-related gamma, alpha, theta and delta oscillations in the EEG? Neuroscience letters, 259(3), 165-168.
[37] Tognoli, E., Kelso, J. A. S. (2015). The coordination dynamics of social neuromarkers. Frontiers in human neuroscience, 9, 563.
[38] von der Malsburg, C., Phillips, W.A., Singer, W. Eds. (2010). Dynamic Coordination in the Brain: From Neurons to Mind, Strüngmann Forum Report, vol. 5. Cambridge, MA: MIT Press.
[39] Tognoli, E., Lagarde, J., de Guzman, G. C., Kelso, J. A. S. (2007). The phi complex as a neuromarker of human social coordination. Proceedings of the National Academy of Sciences, 104(19), 8190-8195.
[40] Tognoli, E., Kelso, J. A. S. (2013a). Spectral dissociation of lateralized pairs of brain rhythms. arXiv preprint arXiv:1310.7662.
[41] Pfurtscheller, G., Stancak, A., Edlinger, G. (1997). On the existence of different types of central beta rhythms below 30 Hz. Electroencephalography and clinical neurophysiology, 102(4), 316-325.
[42] Hari, R. (2006). Action–perception connection and the cortical mu rhythm. Progress in brain research, 159, 253-260.
[43] Jensen, O., Colgin, L. L. (2007). Cross-frequency coupling between neuronal oscillations. Trends in cognitive sciences, 11(7), 267-269.
[44] Assisi, C.G., Jirsa, V.K., Kelso, J.A.S. (2005) Dynamics of multifrequency coordination using parametric driving: Theory and Experiment. Biological Cybernetics, 93, 6-21.
[45] Lehmann, D., Brandeis, D., Ozaki, H., Pal, I. (1987). Human brain EEG fields: micro-states and their functional significance. In Computational Systems—Natural and Artificial (pp. 65-73). Springer Berlin Heidelberg.
[46] Kropff, E., Treves, A. (2007). The complexity of latching transitions in large scale cortical networks. Natural Computing, 6(2), 169-185.
[47] Tsuda, I. (2015). Chaotic itinerancy and its roles in cognitive neurodynamics. Current opinion in neurobiology, 31, 67-71.
[48] Kelso, J.A.S. Fuchs, A. (1995). Self organizing dynamics of the human brain: Critical instabilities and Sil'nikov chaos. Chaos, 5, (1), 64 69.
[49] Benites, D., Tognoli, E., de Guzman, G.C., Kelso, J.A.S. (2010). Brain coordination dynamics: Continuous EEG tracking of the neural functional organization in a social task. Psychophysiology 47, S75–S75.
[50] Benites, D., Tognoli, E., Kelso, J.A.S. (in press). Dinâmicas de Coordenação e Metaestabilidade. In V.G. Haase and G. Gauer (Eds.), Elementos de Psicologia Cognitiva. Porto Alegre: ARTMED.
[51] Fuchs, A., Tognoli, E., Benites, D., Kelso, J.A.S (2010) Neural correlates of social coordination: Spatiotemporal analysis of brain and behavioral measures. In: Society for Neuroscience, 40th meeting, vol. 293, p. 17.
[52] Nadasdy, Z., Benites, D., Shen, J., Briggs, D. E., Lee, M. R., Clarke, D. F., Buchanan, R. J. (2012). Phase divergence of gamma band ECoG predicts epileptiform activity. Program No. 010.09 Neuroscience Meeting Planner. New Orleans, LA.
[53] Lotte, F., Congedo, M., Lécuyer, A., Lamarche, F. (2007). A review of classification algorithms for EEG-based brain–computer interfaces. Journal of neural engineering, 4(2):R1-R13.
[54] Schindler, K., Leung, H., Elger, C. E., Lehnertz, K. (2007). Assessing seizure dynamics by analysing the correlation structure of multichannel intracranial EEG. Brain, 130(1), 65-77.
[55] Tognoli, E., Kelso, J. A. S. (2013b). On the brain's dynamical complexity: coupling and causal influences across spatiotemporal scales. In Advances in Cognitive Neurodynamics (III) (pp. 259-265). Springer Netherlands.
[56] Freeman, W. J. (2006). A cinematographic hypothesis of cortical dynamics in perception. International journal of psychophysiology, 60(2), 149-161.
[57] Bechtel, W. (2004). The epistemology of evidence in cognitive neuroscience. Philosophy and the life sciences: A reader. In R. Skipper Jr., C. Allen, R. A. Ankeny, C. F. Craver, L. Darden, G. Mikkelson, and R. Richardson (eds.), Philosophy and the Life Sciences: A Reader. Cambridge, MA: MIT Press.
[58] McClelland, J. L., Rumelhart, D. E., and the PDP Research Group. (1986). Parallel distributed processing. Explorations in the microstructure of cognition, Psychological and Biological Models, Vol. 2, 216-271.
[59] Mesulam, M. M. (1990). Large-scale neurocognitive networks and distributed processing for attention, language, and memory. Annals of neurology, (28), 597-613.

[60] Canolty, R. T., Knight, R. T. (2010). The functional role of cross-frequency coupling. Trends in cognitive sciences, 14(11), 506-515.
[61] Kelso, J.A.S., Tuller, B. (1984). A dynamical basis for action systems. In M.S. Gazzaniga (Ed.). Handbook of Cognitive Neuroscience (pp. 321 356). New York: Plenum.
[62] Edelman, G. M., Gally, J. A. (2001). Degeneracy and complexity in biological systems. Proceedings of the National Academy of Sciences, 98(24), 13763-13768.
[63] Dodel, S., Tognoli, E., & Kelso, J. A. S. (2013, July). The geometry of behavioral and brain dynamics in team coordination. In International Conference on Augmented Cognition (pp. 133-142). Springer, Berlin, Heidelberg.
[64] Elbert T, Keil A. (2000). Cognitive neuroscience: imaging in the fourth dimension. Nature, 404: 29-31.
[65] Dalal, S. S., Guggisberg, A. G., Edwards, E., Sekihara, K., Findlay, A. M., Canolty, R. T., Berger, M. S., Knight, R. T., Barbaro, N. M., Kirsch, H. E., Nagarajan, S. S. (2008). Five-dimensional neuroimaging: localization of the time–frequency dynamics of cortical activity. Neuroimage, 40(4), 1686-1700002E
[66] Tognoli, E., Kelso, J. A. S. (2010). System and method for analysis of spatio-temporal data. Wipo publication, WO2010016992.